\documentclass[twocolumn,showpacs,preprintnumbers,amsmath,amssymb,prb]{revtex4}

 \def\XXint#1#2#3{{\setbox0=\hbox{$#1{#2#3}{\int}$}
     \vcenter{\hbox{$#2#3$}}\kern-.5\wd0}}

\def\fakebold#1{\relax\ifvmode\leavevmode\fi%
\ifmmode%
\setbox0=\hbox{$#1$}%
\else%
\setbox0=\hbox{#1}%
\fi%
\kern-.02em\copy0 \kern-\wd0%
\kern .04em\copy0 \kern-\wd0%
\kern-.0125em\raise.02em\box0%
}%

\usepackage{graphicx} \usepackage{dcolumn} \usepackage{bm}

\begin{document}



\title{Density dependent spin susceptibility and effective mass in
  interacting quasi-two dimensional electron systems}

\author{Ying Zhang} 
\author{S. Das Sarma} 
\affiliation{Condensed
  Matter Theory Center, Department of Physics, University of Maryland,
  College Park, MD 20742-4111}

\date{\today}

\begin{abstract}
Motivated by recent experimental reports, we carry out a Fermi liquid
many-body calculation of the interaction induced renormalization of
the spin susceptibility and effective mass in realistic two
dimensional (2D) electron systems as a function of carrier density
using the leading-order `ladder-bubble' expansion in the dynamically
screened Coulomb interaction. Using realistic material parameters for
various semiconductor-based 2D systems, we find reasonable
quantitative agreement with recent experimental susceptibility and
effective mass measurements. We point out a number of open questions
regarding quantitative aspects of the comparison between theory and
experiment in low-density 2D electron systems.

\end{abstract}

\pacs{71.10.-w; 71.10.Ca; 73.20.Mf; 73.40.-c}

\maketitle


\section{Introduction}

It is well-known that the mutual Coulomb interaction between electrons
could cause substantial quantitative modification of thermodynamic
properties (e.g. effective mass, specific heat, compressibility,
magnetic susceptibility) in an interacting electron liquid. This is
the so-called many-body renormalization of the Fermi liquid
parameters, which has been studied extensively in three dimensional
metals~\cite{hedin} and in two dimensional (2D) semiconductor
structures~\cite{ando} for a very long time. At zero temperature (or
more generally at low temperatures, $T/T_F \ll 1$ where $T_F = E_F /
k_B$ is the Fermi temperature) the many-body Fermi liquid
renormalization for a quantum electronic system is entirely determined
by the electron density ($n$) with the dimensionless density parameter
$r_s$ being defined as the average inter-electron separation measured
in the units of Bohr radius: $r_s \equiv (\pi n)^{-1/2} / a_B$ where
$a_B = \kappa \hbar^2 / (m e^2)$ is the effective Bohr radius for a
background (lattice) dielectric constant $\kappa$ and a band mass $m$
-- this definition of $r_s$ applies to 2D (in 3D: $r_s \propto
n^{-1/3}$) with $n$ being the 2D electron density. It is easy to see
that our $r_s$-parameter is proportional to the ratio of the average
Coulomb potential energy (i.e.  the interaction energy) to the
noninteracting kinetic energy, and as such the system is strongly
interacting at large $r_s$ (low density) and weakly interacting at
small $r_s$ (high density).  We emphasize that our definition of $r_s$
does not depend on the spin (or valley degeneracy in 2D) of the
system. Studying (and comparing theory with experiment) the density
dependence of various Fermi liquid parameters in interacting electron
liquids has been one of the most important and active many-body
research areas in condensed matter physics in 3D metallic systems (as
well as in normal He-3, a quintessential Fermi liquid albeit with {\em
  a short-range} inter-particle interaction) and more recently, in 2D
electron systems confined in semiconductor structures. The 2D systems
have the distinct advantage of the density being a tunable parameter
so that the density dependence of the Fermi liquid renormalization can
be studied directly. In this paper, we theoretically consider the
density-dependent many-body renormalization of the 2D electronic spin
susceptibility and effective mass, a subject of considerable recent
experimental activity~\cite{okamoto, shashkin, pudalov, tutuc, zhu,
  vakili, shkolnikov, shashkin2, tan} in a number of different
semiconductor heterostructures with confined 2D electron systems.

There has been a number of experimental papers appearing in the recent
literature reporting the low-temperature ($\lesssim 100$mK)
measurement of the susceptibility~\cite{okamoto, shashkin, pudalov,
  tutuc, zhu, vakili, shkolnikov} and effective mass~\cite{shashkin2,
  tan} in 2D electron systems as a function of carrier density in the
$r_s \approx 1 - 10$ parameter range.  Although some aspects of the
data in different experiments (and more importantly, the
interpretation of the data) have been controversial -- most especially
on the issue of whether there is a spontaneous density-driven
ferromagnetic spin polarization transition at low carrier densities in
2D systems -- the experimental reports convincingly establish a strong
enhancement in both the spin susceptibility and effective mass as a
function of decreasing (increasing) carrier density $n$ (interaction
parameter $r_s$). This strong enhancement of the susceptibility with
decreasing carrier density has been demonstrated in 2D electron
systems confined in (100) Si inversion layers, in GaAs
heterostructures, and in AlAs quantum wells. The typical enhancement
in the low temperature susceptibility is in the range of a factor of
$1-4$ for $r_s \approx 1-10$. In addition to the strong low-density
enhancement of the measured low-temperature susceptibility, there are
several other interesting and intriguing features in the experiments.
The susceptibility enhancement shows a modest dependence on the
induced spin polarization (or equivalently, a change in the spin
degeneracy) in the 2D system -- since the susceptibility is typically
measured~\cite{fang} by applying an external magnetic field to produce
a spin splitting in the 2D system, the susceptibility is invariably
measured in the presence of finite spin polarization (and the
extrapolation to zero spin polarization may not always be uniquely
reliable). Another interesting observed recent feature, reported in
AlAs quantum well systems, is that the susceptibility enhancement is
{\em larger} in the single-valley semiconductor system rather than in
the multi-valley system in contrast to the expectation based on just
exchange energy considerations (since the multi-valley system is
naively expected to be more ``dilute'' as the electrons are being
shared among different valleys). The enhancement of effective mass
with decreasing carrier density was observed in Si inversion layer and
GaAs heterostructures. This measurement is performed by examining the
temperature dependence of the low-temperature Shubnikov de Haas
oscillation magnitudes. The effective mass measurement turns out to be
more difficult in general. Very recently, a detailed and extremely
careful experimental measurement of density dependent effective mass,
using several internal consistency checks, has appeared in the
literature~\cite{tan}.

In this paper, we provide an excellent qualitative and reasonable
quantitative (realistic) theoretical understanding of the
density-dependent 2D spin susceptibility and effective mass
measurements at low temperature. Our work, in contrast to much of the
existing theoretical literature on the topic, fully incorporates the
realistic {\em quasi}-2D nature of the electron systems (i.e. the fact
that these systems have finite widths in the transverse direction
normal to the 2D plane of confinement, which considerably modify the
Coulomb interaction between the electrons) which is of substantial
quantitative importance in the experimental parameter regime. We also
study the spin- and the valley-degeneracy dependence of the calculated
2D susceptibility, obtaining in the process qualitative agreement with
the recent experimental finding on the anomalous valley dependence of
the susceptibility in 2D AlAs quantum well structures.  (The spin,
$g_s$ and the valley degeneracy, $g_v$ dependence of the
susceptibility enters through the 2D density of state, which is
proportional to $g_sg_v$.) Our best quantitative agreement with the
experimentally measured 2D effective mass is obtained for the
so-called ``on-shell'' self-energy approximation.

In Sec.~\ref{sec:form} we lay out the theory and formalism of our
calculation. We present the calculated results for spin susceptibility
in Sec.~\ref{sec:sus} and effective mass in Sec.~\ref{sec:mass} for
different realistic 2D systems. In the end in Sec.~\ref{sec:diss} we
discuss upon various issues related to our results and their relations
with the experiments.


\section{Formalism}
\label{sec:form}

We calculate the $T=0$ (paramagnetic) spin susceptibility $\chi^*$ and
effective mass $m^*$ of a {\em quasi}-2D electron system by using the
many-body perturbation theory technique. It has been known~\cite{rice}
for a long time that for an electron system interacting via the
Coulomb interaction, the most important terms are associated with the
long-range divergence (the so-called `ring' or `bubble'
diagrams~\cite{rice}) of the Coulomb interaction, and as such an
expansion in the dynamically screened Coulomb interaction (with the
screening implemented by the infinite series of `bubble' polarization
diagrams) is the appropriate theoretical framework. Such an expansion
is, in fact, asymptotically exact in the $r_s \to 0$ high-density
limit, and is known to work well qualitatively for $r_s > 1$ although
its precise regime of validity can not be determined theoretically and
has to be ascertained by comparing with experiments. The theory
becomes progressively quantitatively worse as $r_s$ increases, but
unless a quantum phase transition intervenes, there is no specific
$r_s$ value which necessarily limits the qualitative validity of the
theory. Motivated by the encouraging fact that this leading-order
expansion in the dynamically screened interaction (involving an
infinite resummation of all the bubble diagrams in the electron
self-energy calculation) seems to provide a good quantitative
description~\cite{rice} for the thermodynamic properties of the 3D
simple metals ($r_s \approx 3-6$), we apply here the same theory for
interacting electrons in {\em quasi}-2D semiconductor structures.

\begin{figure}[htbp]
  \centering
  \includegraphics[width=3in]{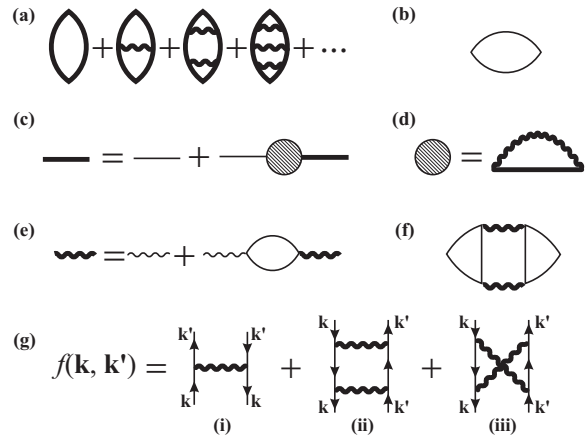}
  \caption{(Color online.) (a) The RPA ladder-bubble series for the 
    interacting susceptibility with the bold straight line the
    interacting Green's function and the bold wavy line the
    dynamically screened interaction; (b) the noninteracting
    susceptibility; (c) the Dyson's equation for the interacting
    Green's function in terms of the noninteracting Green's function
    and the self-energy; (d) the self-energy in the leading-order
    expansion in the dynamical screening; (e) the Dyson's equation for
    the dynamically screened interaction in terms of the bare Coulomb
    interaction (thin wavy lines) and the polarization bubble; (f) a
    charge fluctuation diagram which does not contribute to spin
    susceptibility; (g) Landau's interaction function.}
  \label{fig:feynman}
\end{figure}

Our many-body diagrams for the interacting susceptibility is the
so-called ``ladder-bubble'' series as shown in Fig.~\ref{fig:feynman}
-- this is a consistent conserving approximation for the
susceptibility. Direct calculation of these diagrams turns out to be
difficult for the long ranged Coulomb interaction. However, at $T=0$,
Landau showed that $\chi^*$ can be equivalently expressed through the
Landau's interaction function $f({\bf k}, {\bf k'})$ (shown in
Fig.~\ref{fig:feynman}(g)) as~\cite{rice}:
\begin{equation}
\label{eq:chi}
{\chi \over \chi^*}={m \over m^*} + {1 \over (2 \pi)^2} 
\int f_e (\theta) d \theta,
\end{equation}
where $\chi$ is the Pauli spin susceptibility, $f_e(\theta)$ =
$f_e({\bf k}, {\bf k'})$ with ${\bf k}$ and ${\bf k'}$ on-shell (i.e.
${\bf k}^2/2m = {\bf k'}^2/2m (= E_F)$) is the exchange part of
Landau's interaction function (Fig.~\ref{fig:feynman}(g)(i)), $\theta$
is the angle between ${\bf k}$ and ${\bf k'}$. Similarly, the Landau
theory expression for the effective mass $m^*$ is~\cite{rice}
\begin{equation}
\label{eq:m}
{m \over m^*}=1 - {1 \over (2 \pi)^2} \int f(\theta) d \theta.
\end{equation}
Note that the spin independent exchange Landau's interaction function
$f_e({\bf k}, {\bf k'})$ is responsible for the difference between the
ratio $\chi/\chi^*$ and $m/m^*$~\cite{rice}.

An equivalent, and easier way to derive the effective mass is through
calculating quasiparticle self-energy and obtaining its momentum
derivative. The self-energy within RPA can be written as~\cite{mass}
\begin{equation}
\label{eq:E0}
\Sigma({\bf k}, \omega) = - \int {d^2 q \over (2 \pi)^2} 
\int {d \nu \over 2 \pi i} {v_q \over \epsilon({\bf q}, \nu)} 
G_0 ({\bf q} + {\bf k}, \nu + \omega), 
\end{equation}
where $v_q = F(q) 2 \pi e^2/q$ is the bare electron-electron
interaction and $F(q)$ the {\em quasi}-2D form factor for the
electron-electron interaction~\cite{ando} which will be described in
detail later. Note that this form factor also appears in the
expression of dynamical dielectric function $\epsilon({\bf q}, \nu)$
through $v_q$. As mentioned earlier, a significant quantitative
feature of our theory is the inclusion of the realistic {\em quasi}-2D
Coulomb interaction~\cite{ando} in our calculation which substantially
reduces (compared with the pure 2D case) the quantitative many-body
renormalization effects. In Eq.~(\ref{eq:E0})
\begin{equation}
\label{eq:G0}
G_0({\bf k}, \omega) =
{1 - n_F(\xi_{\bf k}) \over \omega - \xi_{\bf k} + i 0^+ } +
{n_F(\xi_{\bf k}) \over \omega - \xi_{\bf k} - i 0^+ } 
\end{equation}
is the bare Green's function with $n_F$ the Fermi distribution
function and $\xi_{\bf k} = k^2 /(2m) - E_F$ with $E_F$ the
non-interacting Fermi energy. It is shown~\cite{mass} that the
integration along real axis in the expression of self-energy
(Eq.~(\ref{eq:E0})) can be deformed onto imaginary axis, which avoids
the singularities along the real axis and makes the integration
easier.  The contour deformation also breaks the expression of
self-energy into separate terms that correspond respectively to
contributions from the spin-dependent and spin-independent part of the
Landau's interaction function shown in Fig.~\ref{fig:feynman}, and is
very useful for us to derive the expression for susceptibility as
shown below. The expression of the real part of the self-energy can
then be written as
\begin{eqnarray}
\label{eq:ReE}
&& \!\!\!\!\!\!\!\!
\mbox{Re}~\Sigma({\bf k}, \omega) =- \int {d^2 q \over (2 \pi)^2} v_q 
\Theta(2 m \omega + k_F^2 - |{\bf q - k}|^2) \nonumber \\
&& + \int {d^2 q \over (2 \pi)^2} 
v_q \mbox{Re}~{1 \over \epsilon({\bf q}, \xi_{\bf q - k} - \omega)} 
\nonumber \\
&&~~~\times \Big[\Theta(2 m \omega + k_F^2 - |{\bf q - k}|^2) 
- \Theta(k_F^2 - |{\bf q - k}|^2) \Big]\nonumber \\
&& - \int {d^2 q \over (2 \pi)^2} \int {d \nu \over 2 \pi}
v_q \left[ {1 \over \epsilon({\bf q}, i \nu)} -1 \right]
{1 \over i \nu + \omega - \xi_{{\bf q} + {\bf k}}},
\end{eqnarray}
where $k_F$ is the Fermi momentum. The effective mass is derived from
the expression of the real part of the quasiparticle self-energy by
$m/m^* = 1 + (m/k_F) {d \over d k} \mbox{Re}~\Sigma({\bf k}, \xi_{\bf
  k}) |_{k = k_F}$~\cite{rice}.  Combining this with
Eq.~(\ref{eq:chi}), we have
\begin{equation}
\label{eq:chi2}
{\chi \over \chi^*}=1 + {1 \over (2\pi)^2} \int f_e (\theta)
d \theta
+ {m \over k_F} {d \over d k} 
\mbox{Re}~\Sigma({\bf k}, \xi_{\bf k}) |_{k =
  k_F}.
\end{equation}
It is not difficult to show that the second term of Eq.~(\ref{eq:ReE})
accounts for the contribution from the spin-independent exchange
Landau's interaction function $f_e({\bf k}, {\bf k'})$
(Fig.~\ref{fig:feynman}(g)(i)), and therefore the term
$(2\pi)^{-2}\int f_e (\theta) d \theta$ in
Eq.~(\ref{eq:chi2}) exactly cancels the momentum derivative of the
second term in the self-energy Eq.~(\ref{eq:ReE}). Hence the
expression of $\chi / \chi^*$ only contains contributions from the $k$
derivatives of the first and third term in Eq.~(\ref{eq:ReE}). After
converting all the expressions in terms of the dimensionless parameter
$r_s$, and using $2k_F$, $4 E_F$, $2m$ as the momentum, energy, and
mass units, the expression for $\chi^* / \chi \equiv g^* m^* / (g m)$,
where $\chi^* (\chi)$, $g^* (g)$, $m^* (m)$ are respectively the
interacting (noninteracting) spin susceptibility, the Landau
$g$-factor, and the effective mass, is given in our theory as
\begin{eqnarray}
\label{eq:sus}
{\chi \over \chi^*} &=& 
- {2 \alpha r_s \over \pi} \int_0^1 d x {x F(x) \over \sqrt{1 - x^2}}
\nonumber \\
&& + { \sqrt{2} \alpha r_s \over \pi} 
\int_0^\infty x^2 F(x) d x \int_0^\infty d u 
\left[{1 \over \epsilon(x,  iu)} -1 \right] 
\nonumber \\ 
&& ~~~~ \cdot \left[A \sqrt{1 + A/R} - B \sqrt{1-A/R}\right] R^{-5/2},
\end{eqnarray}
where $\alpha = \sqrt{g_v g_s /4}$ with $g_v$ ($g_s$) the valley (spin
degeneracy); $A = x^4 - x^2 - u, B = 2 x u, R = \sqrt{A^2 + B^2}$; $x
= q/(2 k_F)$ and $u = \omega / (4 E_F)$; $\epsilon(x, iu) = 1 + \alpha
r_s g_v g_s F(x) [1/(2x) - \sqrt{A + R} / (2^{3/2} x^3) ]$ is the
imaginary frequency dielectric function. For 2D quantum well, the form
factor $F(q)$ is given by
\begin{eqnarray}
\label{eq:fW}
F(q) = {8 \over q^2d^2 + 4 \pi^2} \left[{3 qd \over 8} 
+ {\pi^2 \over qd} - {4 \pi^4 (1 - e^{-qd}) 
\over q^2d^2 (q^2d^2 + 4 \pi^2)} \right],
\end{eqnarray}
where $d$ is the width of the infinite square-well potential of the
quasi-2D system. For heterostructure quasi-2D systems (e.g. Si
MOSFETs), the form factor is
\begin{equation}
\label{eq:fT}
F(q) = (1 + {\kappa_{\rm ins} \over \kappa_{\rm sc}}) 
{8 + 9qb + 3q^2b^2 \over 8 (1+qb)^3 }
+ (1 - {\kappa_{\rm ins} \over \kappa_{\rm sc}}) {1 \over 2 (1 + qb)^6},
\end{equation}
with $b = \left[ \kappa_{\rm sc} \hbar^2 /( 48 \pi m_z e^2 n^*)
\right]^{1/3}$ defining the width of the quasi 2D electron gas,
$\kappa_{\rm sc}$ and $\kappa_{\rm ins}$ are the dielectric constants
for the space charge layer and the insulator layer, $m_z$ is the band
mass in the direction perpendicular to the quasi 2D layer, and $n^* =
n_{\rm depl} + {11 \over 32}n$ with $n_{\rm depl}$ the depletion layer
charge density and $n$ the 2D electron density. We choose $n_{\rm
  depl}$ to be zero in our calculations, since it is unknown in
general (finite small value of $n_{\rm depl}$ do not change our
results). From Eqs.~(\ref{eq:chi}) and (\ref{eq:m}) we have
\begin{equation}
\label{eq:mr}
{m \over m^*} = {\chi \over \chi^*} 
+ {\alpha r_s \over \pi} \int_0^1 dx {F(x) \over x \epsilon(x,0)}.
\end{equation}

To avoid any possible confusion, we emphasize that the definition of
our spin susceptibility is the derivative of the magnetization with
respect to the applied magnetic field, and therefore $\chi^*$ is
well-defined even in a non-zero parallel magnetic field $B$. It is
important to note and easy to show that, within our approximation in
which the spin-orbital effect can be neglected, the parallel field $B$
dependence of the spin susceptibility $\chi^*$ and the $g$-factor
$g^*$ manifests only through the spin-degeneracy factor $g_s$, which
is present in the expression (\ref{eq:sus}).

Motivated by recent experimental studies, we have directly evaluated
the interacting susceptibility as a function of density at different
spin-($g_s = 1$, $2$) and valley-degeneracy ($g_v = 1$, $2$ for AlAs
quantum wells) for three different 2D semiconductor systems: n-Si(100)
inversion layers; n-GaAs gated undoped heterostructures;
modulation-doped AlAs quantum wells.  We also calculated effective
mass for n-GaAs heterostructures motivated by a very recent
experimental report~\cite{tan}. In the rest of this paper, we present
and discuss our calculated results for $\chi^*/\chi$ and $m^*/m$ in
light of the recent experimental data in 2D semiconductor structures.


\section{Results for spin susceptibility}
\label{sec:sus}

\begin{figure}[htbp]
\centering \includegraphics[width=3in]{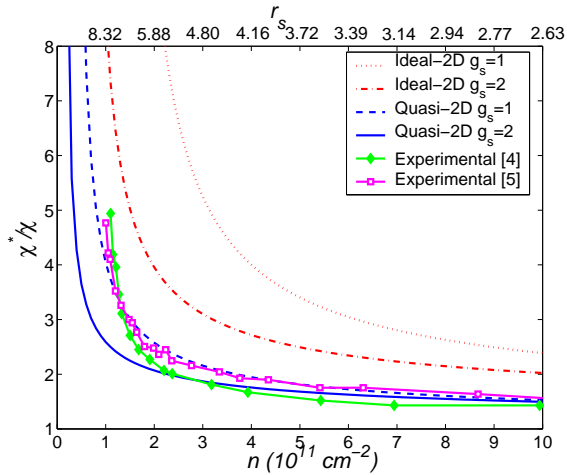}
  \caption{(Color Online.) Calculated spin susceptibility of Si 
    MOSFET ideal and {\em quasi}-2D systems with $g_s = 1$ and $2$, to
    be compared with experimental data from
    Refs.~\onlinecite{shashkin, pudalov}.}
  \label{fig:Si}
\end{figure}

In Fig.~\ref{fig:Si} we show our (100) Si MOSFET results (taking a
valley degeneracy $g_v = 2$) for the calculated susceptibility
comparing with the recent experimental results. We show two sets of
results corresponding to the realistic {\em quasi}-2D system and the
strict 2D system. In each case we show the calculated susceptibility
for both the $g_s = 1$ (due to the presence of an external magnetic
field) and the normal $g_s = 2$ situation. Since the experiments are
invariably carried out in the presence of finite external magnetic
fields, the experimental results probably correspond to the region in
between the $g_s = 1$ and $g_s = 2$ theoretical curves (for the {\em
  quasi}-2D system). A more detailed discussion is made in
Sec.~\ref{sec:diss} on the reason why we are comparing the
experimental data with our $g_s = 1$ and $g_s = 2$ theoretical
results. There are three important points we make about
Fig.~\ref{fig:Si}: (1) The {\em quasi}-2D results are lower than the
2D results by a factor of $1.5$ to $3$, and the relative difference is
much larger at low carrier densities since the effective {\em
  quasi}-2D layer width is larger at lower 2D densities; (2) the
theory gives a reasonable description of the experimental data, -- in
particular, the experimental data points lie very close to the region
between the theoretical $g_s = 1$ and $g_s = 2$ {\em quasi}-2D
susceptibility results; (3) the pure-2D results disagree with
experiments.

\begin{figure}[htbp]
  \centering
  \includegraphics[width=3in]{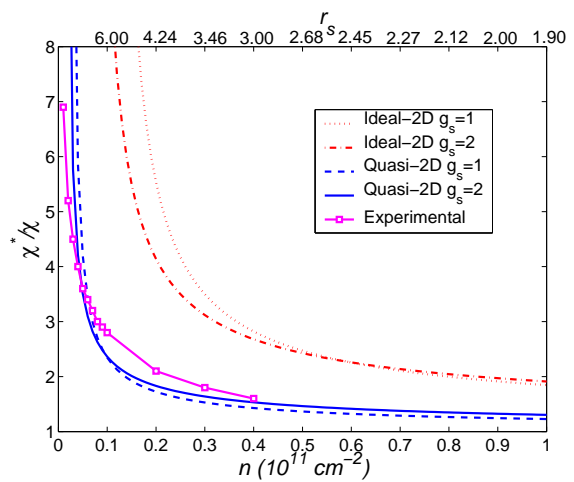}
  \caption{(Color online.) Calculated spin susceptibility of GaAs
    heterostructure ideal and {\em quasi}-2D systems with $g_s = 1$
    and $2$, to be compared with experimental data from
    Refs.~\onlinecite{zhu}.}
  \label{fig:Ga}
\end{figure}

In Fig.~\ref{fig:Ga} we show our theoretical susceptibility results
for electrons confined in {\em quasi}-2D GaAs heterostructure (valley
degeneracy $g_v = 1$) comparing with the recent measurements of Zhu
{\it et. al.}~\cite{zhu}. Again, the agreement between our results and
the experimental measurements is very good for the realistic {\em
  quasi}-2D calculations. Note that at the low carrier densities of
interest in the GaAs 2D system, the {\em quasi}-2D effects are
extremely strong (almost a factor of $5$!), and the quantum Monte
Carlo (QMC) calculation~\cite{attaccalite} with which the experimental
results were compared in Ref.~\onlinecite{zhu} are completely
inapplicable since they are for a strict 2D system rather than a {\em
  quasi}-2D system. In fact, any agreement between the measured
susceptibility and the QMC results for the ideal zero-width 2D system
should be taken as a spectacular quantitative failure for the QMC
theory in the low-density regime. The reason for such a strong {\em
  quasi}-2D effect in GaAs is that at very low densities ($n \sim 10^9
\mbox{cm}^{-2}$), the transverse {\em quasi}-2D width of the electron
wavefunction is extremely large ($> 500 \AA$) so that the Coulomb
interaction is substantially suppressed compared with the strict 2D
limit result.  (This demonstrates that it is misleading to compare
{\em quasi}-2D experimental results with the strict 2D theory as is
often done in the literature!).

\begin{figure}[htbp]
  \centering
  \includegraphics[width=3in]{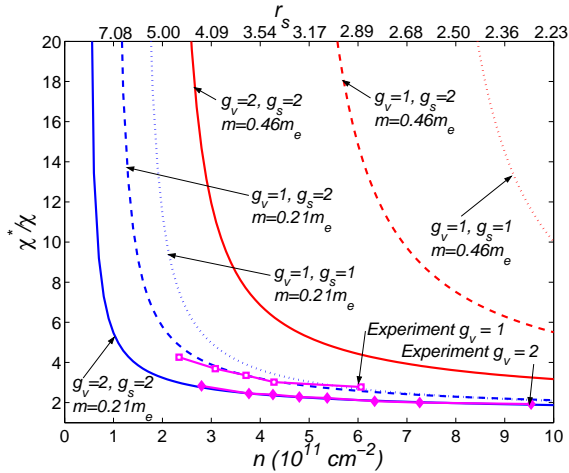}
  \caption{(Color online.) Calculated spin susceptibility of AlAs 
    quantum well ideal and {\em quasi}-2D systems with $g_s = 1$ and
    $2$ and $g_v = 1$ and $2$, to be compared with experimental data
    from Refs.~\onlinecite{shkolnikov}. Effective $\kappa = 10$. We
    use two different values of band mass: $m = 0.21 m_e$ and $m =
    0.46 m_e$.}
  \label{fig:Al}
\end{figure}

In Fig.~\ref{fig:Al} we show our theoretical results for AlAs quantum
wells. The AlAs quantum wells used in Ref.~\onlinecite{vakili,
  shkolnikov} are rather narrow (width $\sim 50 \AA$), and the
additional self-consistent confining potential arising from the 2D
electrons themselves produces even stronger confinement of the
carriers thus further narrowing the effective {\em quasi}-2D width.
Therefore, we show only the strict-2D results for the 2D AlAs system
in Fig.~\ref{fig:Al} for the sake of clarity. Our corresponding {\em
  quasi}-2D results are somewhat below the theoretical curves for the
strict-2D case. In Fig.~\ref{fig:Al} we show our theoretical results
for two different valley degeneracy ($g_v = 1$ and $2$) situations.
Our theoretical investigation of the valley degeneracy dependence of
the 2D spin susceptibility behavior is necessitated by the puzzling
recent experimental finding~\cite{vakili,shkolnikov} of an interesting
valley degeneracy dependence in AlAs systems, namely, the many-body
enhancement for $\chi^* / \chi$ is larger for the valley occupancy of
$1$ than the valley occupancy of $2$. (Our results for Si MOSFETs and
GaAs heterostructures in Figs. 1 and 2 are for valley degeneracy
values $g_v = 2$ and $1$ respectively, consistent with the known band
structure for Si and GaAs.)

In Fig.~\ref{fig:Al}, the results for $m = 0.21m_e$, which corresponds
to the transverse AlAs band mass $m_t$, are in reasonable quantitative
agreement with the experimental AlAs quantum well results, including
the anomalous finding of $\chi^* / \chi$ having a stronger many-body
Fermi liquid enhancement for the lower valley occupancy of $1$ than
for the higher occupancy of $2$. We note that at high enough electron
densities ($\gtrsim 10^{12} \mbox{cm}^{-2}$), this anomalous valley
dependence of the spin susceptibility would disappear according to our
theoretical calculations with $\chi^*/\chi$ for $g_v = 2$ being larger
than that for $g_v = 1$, consistent with one's naive expectations. It
may be more appropriate to use~\cite{shayegan} a larger 2D band mass
in our calculation given by $m = 0.46 m_e = \sqrt{m_t m_l}$ with $m_l
= m_e$ the longitudinal band mass. Use of this band mass results in
larger theoretical values for spin susceptibility than that was
measured in experiment, but the trend and the characteristic of
valley-degeneracy dependence remains the same. The exact cause of the
quantitative difference between our $m = 0.46m_e$ results and the
experiments probably lies in the yet unknown details of the samples
and experimental procedures, and are therefore not clear to us right
now.

Why does the valley occupancy dependence of the many-body
susceptibility enhancement act in such a way? The reason lies in
many-body correlation effects beyond the naive exchange energy
considerations.  Within an exchange-only Hartree-Fock theory, in fact,
the $2$-valley occupancy state would have a higher susceptibility
enhancement than the $1$-valley state. But correlation effects are
important in the system, and at low enough carrier densities the
$1$-valley state turns out to have stronger many-body effects than the
$2$-valley state. Our theory, which is essentially a self-consistent
dynamically screened Hartree-Fock theory, includes correlation effects
demonstrating that at low enough carrier densities, $\chi^* / \chi$
could be enhanced in the $1$-valley state over the $2$-valley state as
has been experimentally observed. Another way to understand this is
through the screening effect. As $g_v$ increases, Fermi momentum
decreases as $k_F \propto 1/\sqrt{g_v}$, which favors many-body
renormalization.  However, the screening effect increases with
increasing $g_v$ since $q_{TF} \propto g_v$, which tends to decrease
the renormalization effect. These two effects are competing with each
other. At low densities, the screening effect is predominant, and
therefore a smaller $g_v$ results in much less screening than a bigger
$g_v$ and hence produces larger susceptibility renormalization. On the
other hand, at high densities the screening effect is less important,
and a smaller $g_v$ results in a larger Fermi energy and a smaller
renormalization effect. In particular, the valley degeneracy
dependence of many-body effects should qualitatively correlate with
the dimensionless parameter $q_{TF}/(2k_F) \propto g_v^{3/2}$: for
large (small) $q_{TF}/(2k_F)$, low (high) density, smaller (larger)
values of $g_v$ produce larger renormalization. Note that the
experimental results in Ref.~\onlinecite{shkolnikov} do not show any
strong $g_s$ dependence in contrast to our theoretical results in
Fig.~\ref{fig:Al}. We discuss this puzzle in Sec.~\ref{sec:diss} below
of our paper.


\section{Results for effective mass}
\label{sec:mass}

\begin{figure}[htbp]
\centering \includegraphics[width=3in]{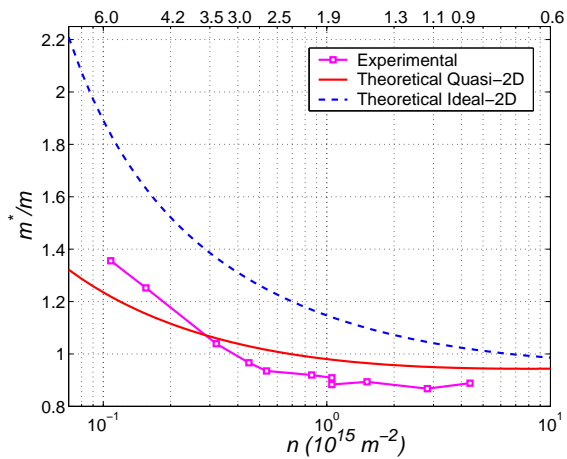}
  \caption{(Color online.) Calculated effective mass of GaAs
    heterostructure ideal and {\em quasi}-2D systems with $g_s = 2$
    and $g_v = 1$, to be compared with experimental data from
    Refs.~\onlinecite{tan}.}
  \label{fig:mass}
\end{figure}

Although the main task of our current work is the calculation of
interacting 2D susceptibility to compare with the recent experimental
results~\cite{okamoto, shashkin, pudalov, tutuc, zhu, vakili,
  shkolnikov, shashkin2}, a very recent experimental report~\cite{tan}
of 2D n-GaAs effective mass measurement allows us to apply our
recently developed theory~\cite{mass} for the quasiparticle effective
mass enhancement to 2D GaAs in order to compare with these
experimental results. For details on the quasiparticle effective mass
theory we refer to our recent publication~\cite{mass}.

In Fig.~\ref{fig:mass} we show our theoretical effective mass results
for electrons confined in {\em quasi}-2D GaAs heterostructure (valley
degeneracy $g_v = 1$) comparing with the recent measurements of Tan
{\it et. al.}~\cite{tan}. The agreement between our results and the
experimental measurements is very good for the realistic {\em
  quasi}-2D calculation. Again the ideal-2D effective mass results
turn out to be much larger than the experimental data, emphasizing
once more the importance of the {\em quasi}-2D effect on the many-body
renormalization of physical quantities in such systems. We emphasize
that, while at high densities the quasi-2D results are rather close
(with a $10-20\%$ difference) to the pure 2D results, at low densities
this difference could be as large as a factor $2$.


\section{Discussion}
\label{sec:diss}

Before concluding we make some critical observations and comments on
our theory and its implications and, more particularly, on the
comparison with the recent experimental results. One of our important
conclusions is that the inclusion of {\em quasi}-2D form factor
effects is essential in understanding the 2D susceptibility results.
As such, an important issue is the realistic nature of our model where
we have included the {\em quasi}-2D form-factor effects through the
standard variational approximation~\cite{ando} for heterostructures.
We believe that our {\em quasi}-2D electronic structure model is
extremely reasonable, but some uncertainty and error probably arise
(particularly at low carrier densities) from our lack of knowledge
about the depletion charge density in these systems.

A rather important factor in the comparison with the experimental data
that is left out of our consideration is the effect of orbital
coupling~\cite{dassarma} of the in-plane component of the external
magnetic field invariably present in the experimental measurement of
the spin susceptibility. At low carrier densities, when the {\em
  quasi}-2D width of the 2D layer is large, such a magneto-orbital
coupling could have substantial effects on the 2D effective
mass~\cite{dassarma}.  This effect is, however, entirely of
one-electron origin, and we assume, somewhat uncritically at this
stage, that the magneto-orbital effect drops out of the susceptibility
enhancement factor $\chi^* / \chi$ since the enhancement involves a
ratio of the interacting and the non-interacting effective mass both
of which will be affected in a similar manner by the magneto-orbital
effect. In any case, the magneto-orbital effect is negligibly small
for Si MOSFETs and AlAs quantum wells because of their tight {\em
  quasi}-2D confinement, and in the n-GaAs structure the {\em ratio}
$\chi^* / \chi$ should not be much affected by the magneto-orbital
coupling. Only at rather low densities in GaAs heterostructures the
magneto-orbital effect may play a role.

\begin{figure}[htbp]
\centering \includegraphics[width=3in]{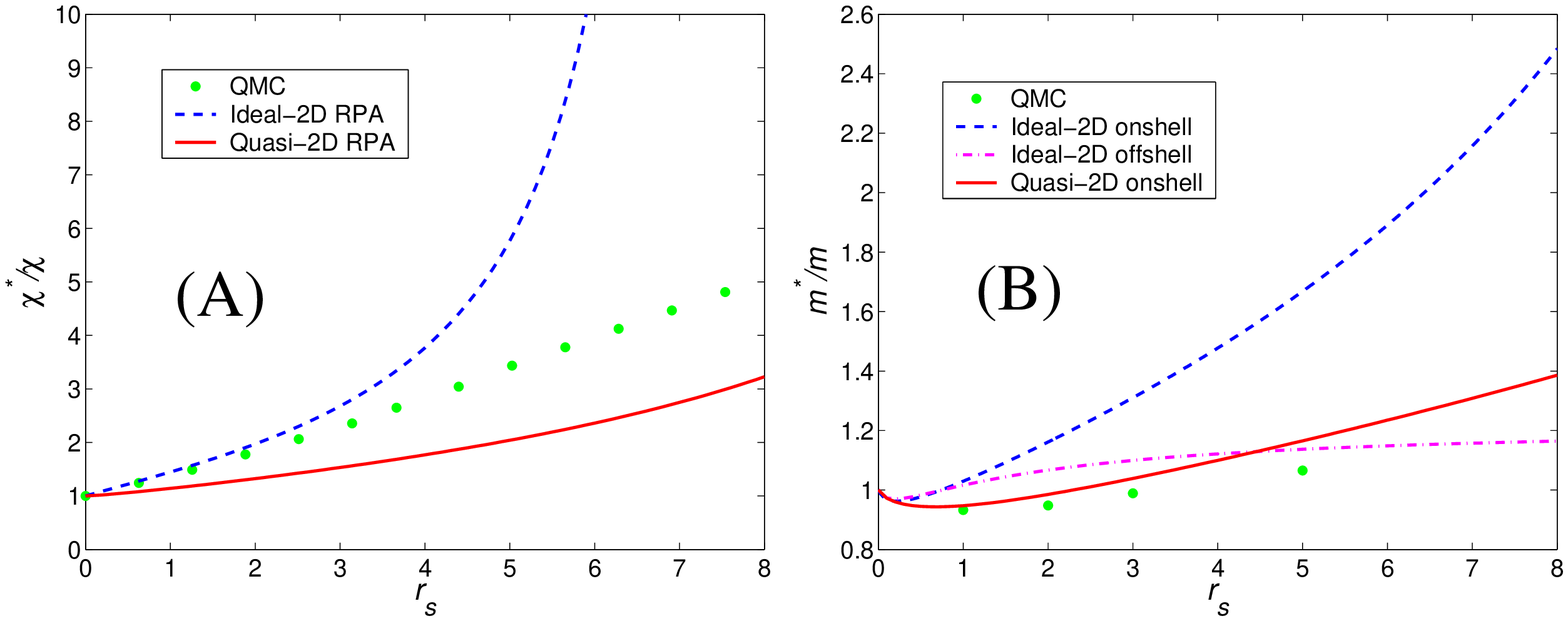}
   \caption{(Color online.) (A) Comparison between the results of 
     spin susceptibility calculated using QMC
     method~\cite{attaccalite} and our results calculated within RPA
     for a 2D electron system with $g_s = 2, g_v = 1$. Our results are
     for both ideal and quasi-2D systems as shown; the QMC results are
     for ideal 2D systems. (B). Comparison between our RPA many-body
     theory and the QMC results~\cite{kwon} for quasiparticle
     effective mass.  Results are shown for strict 2D (both on-shell
     and off-shell RPA self-energy approximations as well as QMC) and
     quasi-2D (on-shell RPA self-energy approximation) cases. Our
     effective mass results shown in Fig. 5 correspond to the on-shell
     approximation which is the consistent approximation for the Fermi
     liquid effective mass within the RPA self-energy approximation.}
  \label{fig:QMC}
\end{figure}

Although our {\em quasi}-2D model is reasonably realistic and
accurate, our many-body spin susceptibility calculation is necessarily
based on approximations since the problem of an interacting quantum
Coulomb system cannot be solved exactly. We emphasize in this respect
that QMC calculations, while being accurate as a matter of principle,
turn out as a matter of practice to be often unreliable due to various
approximations (e.g.  node fixing, back-flow correlations, etc.) and
inaccuracies (e.g.  finite size effects, numerical errors) inherent in
the QMC technique.  Also the QMC calculations leave out {\em quasi}-2D
effects, and are therefore quantitatively incorrect. In
Fig.~\ref{fig:QMC} we make a comparison between our result and the QMC
result~\cite{attaccalite} for susceptibility in an ideal-2D electron
system with $g_s = 2$ and $g_v = 1$. In the intermediate and large
$r_s$ region, the difference between the results of RPA and QMC
methods turn to be very large. But we note that our {\em quasi}-2D
results are quantitatively quite close to the QMC pure-2D results. (A
similar situation seems to hold for the effective mass calculations
also.) This demonstrates the care that is needed in comparing theory
with experiments in this problem.

In comparing theory and experiment for the susceptibility and the
effective mass in 2D electron systems it is important to emphasize the
fact that the experimental measurements for $\chi^*$ and $m^*$ in 2D
systems are typically {\em not} thermodynamic measurements, but are
essentially transport measurements of thermodynamic quantities in an
applied magnetic field which spin-polarizes the system through the
Zeeman splitting. For example, the 2D susceptibility is
measured~\cite{okamoto, shashkin, pudalov, tutuc, zhu, vakili,
  shkolnikov, shashkin2} by either monitoring~\cite{okamoto} the full
spin polarization of a 2D system in an applied parallel field or
through the clever coincidence technique~\cite{fang} in a titled
magnetic field. We want to raise three important issues related to
these experimental techniques and their relevance to our theory.
First, strictly speaking these techniques does not provide a true
measurements of the spin susceptibility $\chi^*$ that is defined
through $\chi^*(B) = \partial M / \partial B$ with $M$ the
magnetization and $B$ the applied magnetic field. In all these
techniques, the experimentally measured quantity is actually $M/B$,
which we now call $\tilde{\chi}^*(B)$. As has been explained in the
experimental literature~\cite{okamoto, shashkin, pudalov, tutuc, zhu,
  vakili, shkolnikov}, these studies measure indirectly the
occupancies of spin up and down levels through transport or Landau
level coincidence measurements. The spin susceptibility is then
inferred from these indirect spin polarization studies by using the
simple non-interacting spin-dependent density formula: In the
transport measurement technique, $g^* \mu_B B_p = E_F$ where $B_p$ is
the so-called saturation (parallel) field for `complete spin
polarization'. In the Landau level coincidence technique, $g^* \mu_B B
= i \omega_c = i e \hbar B_{\perp}/m^*$ with $i$ half integer or
integer. In either case, the derived spin susceptibility is exactly
$\tilde{\chi}^*(B) = M/B$ that we have just discussed. ($B = B_p$ in
the transport measurement technique, and $B$ corresponds to the Landau
level coincidence magnetic field in the Landau level coincidence
technique.)  It is important to note that $\tilde{\chi}^*(B) \ne
\chi^*(B) \ne \chi^*(B=0)$ for $B >> 0$, and these three quantities
coincide in the $B \to 0$ limit. The precise definition of spin
susceptibility was never explicitly made clear in the previous
experimental literature.  Second, even though the experimentally
observed $\tilde{\chi}^*(B)$ is not exactly the thermodynamic quantity
$\chi^* = \chi^*(B=0)$, in normal circumstances they are close to each
other. Especially, if the magnetization curve of the system is smooth
(which normally it will be), $\tilde{\chi}^*(B)$ is in between
$\chi^*(B=0)$ and $\chi^*(B=B_p)$ which are the two extreme case we
are considering in this work. As we mentioned before, since the
magnetic field dependence of $\chi^*$ only manifests itself through
the spin degeneracy factor, we can say that the experimentally
measured $\tilde{\chi}^*(B)$ should be in between $\chi^*(g_s = 2)$
and $\chi^*(g_s = 1)$, which is the justification for our comparison
of the experimental data with our $g_s = 1, 2$ theoretical results.
In Ref.~\onlinecite{zhu}, the issue of finite field and hence the spin
degeneracy factor dependence of the susceptibility is made more clear.
However, in this case not only the partial spin polarization need to
be considered, the effect of the finite perpendicular magnetic field
and Landau levels on the susceptibility need to be further
investigated, which is beyond the scope of this work. Third, the
difference between $\tilde{\chi}^*(B)$ and $\chi^*$ may be helpful in
understanding some of the discrepancies between the experimental data
and our theoretical findings. For example the recent
measurements~\cite{vakili, shkolnikov} in 2D AlAs quantum well systems
find a strongly valley-degeneracy-dependent spin susceptibility which,
however, demonstrates essentially {\em no} spin degeneracy dependence.
Since the valley index is essentially a pseudo-spin index, theory
would predict the {\em same} valley and spin degeneracy dependence of
susceptibility unless there is strong spin-orbit or valley-orbit band
structure effects in the system.  Therefore, in the $r_s$ region where
the spin susceptibility has a large dependence on the valley
degeneracy, the spin degeneracy should play an equally big role, as
shown in Fig.~\ref{fig:Al}. This theory-experimental discrepancy may
very well be a result of the difference between $\tilde{\chi}^*(B)$
and $\chi^*$. This factor might have also caused the quantitative
discrepancy between our RPA results (using $m = 0.46 m_e$) and the
experimental results shown in Fig.~\ref{fig:Al}.  Other possible
explanation could also lie in one-electron band structure physics, not
in many-body theory. There have been no QMC calculations revealing the
spin and valley degeneracy dependency of the spin susceptibility.

The many-body approximation we use in our work, namely the
renormalized interaction calculated within the infinite series of
bubble diagrams and the susceptibility calculated in the infinite
series of ladder diagrams using the appropriate dynamically screened
interaction (i.e. the ladder-bubble approximation for $\chi^*$),
corresponds to the leading-order expansion in the dynamically screened
interaction. Such an expansion is asymptotically exact in the
high-density limit, and is known to work well in the low-density limit
as well. Unfortunately, its regime of quantitative validity is
unknown, and it is likely to become progressively quantitatively
inaccurate as $r_s$ increases. It should, however, be qualitatively
valid for larger $r_s$ values as long as the interacting electron
system remains a Fermi liquid. Our reasonable agreement (without
adjusting any parameters) with existing experiments shows that the
theory remains well-valid in 2D-systems at least up-to $r_s \sim 7$,
which is consistent with the success of the corresponding 3D theory in
the metallic densities ($r_s \sim 3 - 6$). Since no systematic and
uncontrolled many-body theory approximation is available for
susceptibility beyond the ladder-bubble series approximation carried
out in this paper, we can only test the validity of our theory through
the direct comparison with experiments, and on this ground the theory
seems to be well-justified.

Our quasiparticle effective mass results (Fig.~\ref{fig:mass}) are
based on the theory developed by us in Ref.~\onlinecite{mass}
recently. This theory is the RPA theory, based on the leading-order
self-energy calculation in the dynamically screened Coulomb
interaction. While the theory is asymptotically exact in the high
density $r_s \to 0$ limit, it is not based on an $r_s$ expansion and
is essentially a self-consistent field theory. Unfortunately, however,
like many other self-consistent filed theories, the level of validity
of this theory for large $r_s$ (i.e. the strongly interacting regime)
is unknown except that it is expected to remain qualitatively
valid~\cite{hedin, rice, mass} for $r_s > 1$ unless there is a quantum
phase transition. Since we have discussed the approximation scheme and
the validity of our RPA effective mass theory in some details in
Ref.~\onlinecite{mass}, we do not repeat those arguments here. We
point out that the effective mass results given in Fig.~\ref{fig:mass}
use the ``on-shell'' self-energy approximation~\cite{rice, mass},
which has been argued~\cite{rice, mass} to be the correct dynamical
approximation consistent with the Landau Fermi liquid theory as long
as the self-energy is obtained in the leading-order dynamically
screened interaction (i.e. RPA). Our ``off-shell'' approximation for
the 2D effective mass (shown in Fig.~\ref{fig:QMC}(B)) differs from
the corresponding ``on-shell'' results by a factor of $2$ or more at
low densities, and show different trends as well. The fact that the
``on-shell'' effective mass results agree much better with experiment
than the ``off-shell'' results is additional evidence in support of
the ``on-shell'' approximation being the correct one within RPA
self-energy scheme. One can try to ``improve'' upon RPA by including
local field corrections~\cite{marmorkos} to the dynamical electron
polarizability (i.e.  bare bubble of RPA) which, in some crude manner,
simulates the incorporation of higher-order vertex corrections in the
theory. But such local field corrections are uncontrolled, and
probably inconsistent, since many diagrams in the same order are
typically left out. We are therefore unconvinced that the inclusion of
local field corrections in the theory is necessarily an improvement on
RPA. The great conceptual advantage of RPA is that it is a
well-defined approximation that is both highly physically motivated
(i.e. dynamical screening) and theoretically exact in the high-density
($r_s \to 0$) limit. Attempted improvement upon RPA through the
arbitrary inclusion of local field correction may neither be
theoretically justifiable nor more reliable.  Keeping these caveats in
mind, we mention that several previous works~\cite{marmorkos, mass}
have shown that local field corrections to the many-body properties of
two dimensional electron gas turn to be very small in the density
range of our current interest.

In has recently been discussed in detail by us~\cite{mass} that RPA is
not necessarily a high-density theory although it is exact in the high
density limit. The ring diagram approximation, which is at the heart
of RPA, is an expansion in the dynamically screened interaction. Under
some circumstance RPA can be very loosely thought of~\cite{mass,
  hedin} as an expansion in an effective parameter $r_s/(r_s + C)$
where $C > 1$ is a constant. This implies that RPA could in some
situation turn out to be decent approximation in the $r_s > 1$ case.
Indeed 3D metals, with $3<r_s<6$, seem to be reasonably
well-described~\cite{hedin} by RPA. Leading-order vertex corrections
to RPA~\cite{rice, mass, marmorkos} typically produce only small
quantitative corrections in the $r_s > 1$ regime, again empirically
justifying the validity of RPA. There is obviously a lot of
cancellation among the higher-order diagrams (in particular, between
the self-energy and the vertex diagrams) leading to the good empirical
success of RPA consisting only of the bare bubble diagrams.  But this
empirical success is physically well-motivated, arising from the
long-range nature of the inter-electron Coulomb interaction with the
decisive point of physics being the dynamical screening properties of
a Coulomb liquid. In an interacting system with short-range bare
interactions, e.g. neutral He-3, the ring-diagrams play no special
role, and RPA is not a meaningful approximation. In a strongly
interacting system, such as a 2D electron system with $r_s > 1$, one
must use a careful and critical comparison between theory and
experiment as the principal guide in deciding the validity of any
theory. By this standard RPA appears to be a reasonable approximation
even for $r_s > 1$. There are many other theoretical techniques in
condensed matter physics which work well beyond their putative regime
of validity -- for example, the dynamical mean field theory (DMFT),
which is exact in infinite dimensions, seems to provide good results
in 2 and 3 dimensional strongly correlated systems, and the local
density approximation (LDA), which is exact for very slowly varying
density inhomogeneity, works well for the band structure of real
solids with their rapidly varying densities. Similar to DMFT and LDA,
RPA works well because it is fundamentally a non-perturbative theory
which accounts for some key ingredient of the interaction physics,
namely, screening in the situation under consideration in this paper.

We conclude by emphasizing that the most significant implication of
the excellent qualitative and quantitative agreement between our
theory and 2D spin susceptibility and effective mass measurements is
that interacting 2D electron systems, which have been of much interest
recently, remain Fermi liquids down to reasonably low carrier
densities ($r_s \sim 7$ or so at least), and many-body perturbative
techniques, coupled with a realistic {\em quasi}-2D description for
the electronic structure, provide a very good qualitative and
quantitative model for these systems. There is no need to invoke any
non-Fermi liquid concepts to explain the recent experimental results
on quasiparticle renormalization effects~\cite{okamoto, shashkin,
  pudalov, tutuc, zhu, vakili, shkolnikov, shashkin2, tan}.

This work is supported by the NSF, the DARPA, the US-ONR, and the LPS.


\bibliography{sus}

\end{document}